\documentclass[fdp,a4paper,fleqn]{w-art}
\usepackage{times,cite,w-thm}
\usepackage[]{graphicx}

\usepackage{amsmath,amssymb}           
\usepackage{epsfig}                    
\usepackage[matrix,arrow]{xy}          
\usepackage{xspace}                    
\usepackage{stmaryrd}                  
\usepackage{slashed}

\DeclareSymbolFont{bbold}{U}{bbold}{m}{n}
\DeclareSymbolFontAlphabet{\mathbbold}{bbold}

\newcommand{\bq}{\begin{equation}}
\newcommand{\eq}{\end{equation}}
\newcommand{\bea}{\begin{eqnarray}}
\newcommand{\eea}{\end{eqnarray}}

\newcommand{\dd}{\mathrm{d}}
\newcommand{\ee}{\mathrm{e}}

\newcommand{\der}{\partial}

\newcommand{\bbR}{\mathbb{R}}

\DeclareMathOperator{\GL}{\mathit{GL}}

\DeclareMathOperator{\Spin}{\mathit{Spin}}

\DeclareMathOperator{\Cliff}{Cliff}

\DeclareMathOperator{\vol}{vol}

\newcommand{\Gs}[1]{\Gamma(#1)}

\newcommand{\Lgen}{L}

\newcommand{\Dgen}{{D}}

\newcommand{\GM}[2]{\big<#1,#2\big>}

\newcommand{\bisp}[1]{#1_{\!\scriptscriptstyle\#}}

\newcommand{\GenRic}{R^{\scriptscriptstyle 0}}
\newcommand{\GenS}{R}

\newcommand{\Ggeom}{G_{\textrm{split}}}

\newcommand{\mukai}[2]{\big<{#1},{#2}\big>}

\begin{document}

\pagespan{1}{}



\title[Generalised Geometry and type II Supergravity]{Generalised Geometry and type II Supergravity}


\author[A. C.]{Andr\'e Coimbra\inst{1}\footnote{Corresponding author\quad E-mail:~\textsf{a.coimbra08@imperial.ac.uk}}}
\address[\inst{1}]{ Department of Physics, \\Imperial College London \\   London, SW7 2AZ, UK}
\author[C. S-C.]{Charles Strickland-Constable\inst{1}}
\author[D. W.]{Daniel Waldram\inst{1}}
\begin{abstract}
Ten-dimensional type II supergravity can be reformulated as a generalised
geometrical analogue of Einstein gravity, defined by an
$O(9,1)\times O(1,9)\subset O(10,10)\times\bbR^+$ structure on the
generalised tangent space. To leading order in the fermion fields, this allow one to rewrite the action, equations of motion and
supersymmetry variations in a simple, manifestly
$\Spin(9,1)\times\Spin(1,9)$-covariant form.
\end{abstract}
\maketitle                   






\paragraph{Introduction}
\label{sec:intro}


Generalised geometry~\cite{GCY,Gualtieri} is the study of structures
on a generalised tangent space $E\simeq TM\oplus T^*M$. Local
diffeomorphism invariance is replaced by a larger group that also
includes the gauge transformations of the NSNS two-form $B$ and there
is a natural $O(d,d)$ structure on $E$.

In ~\cite{Main} we showed that ten-dimensional type IIA and IIB
supergravity theories, in the context of the ``democratic formalism'' of~\cite{democratic}, can be formulated as generalised geometrical analogues of Einstein gravity. 


\paragraph{Generalised structure bundle}
\label{sec:OddGG}

Let $M$ be a 10-dimensional spin manifold. The generalised tangent space $E$ is
an extension of the tangent space by the cotangent space 
\begin{equation}
\label{eq:Odd-Edef}
   0 \longrightarrow T^*M \longrightarrow E 
      \longrightarrow TM \longrightarrow 0 ,
\end{equation}
which depends on patching one-forms $\Lambda_{(ij)}$. If
$v_{(i)}\in TU_i$ and $\lambda_{(i)}\in T^*U_i$, such that
$V_{(i)}=v_{(i)}+\lambda_{(i)}$ is a section of $E$ over the patch 
$U_i$, then $
   v_{(i)} + \lambda_{(i)} 
      = v_{(j)} + \big( \lambda_{(j)} -
        i_{v_{(j)}}\dd\Lambda_{(ij)}\big)  $ on the overlap $U_i\cap U_j$ and consistency requires the $\Lambda_{(ij)}$ also satisfy higher-order cocycle conditions. 

In order to describe the dilaton correctly we 
consider a slight generalisation of $E$. We define the bundle
\begin{equation}
\label{eq:Ep-def}
   \tilde{E} = \det T^*M \otimes E . 
\end{equation}
$\tilde{E}$ then has a natural $O(10,10)\times\bbR^+$ structure. A
\emph{conformal basis} $\{\hat{E}_A\}$ with $A=1,\dots 20$ on
$\tilde{E}_x$ is one satisfying  $
   \GM{\hat{E}_A}{\hat{E}_B} = \Phi^2 \eta_{AB}$ where $\eta$ is the $O(10,10)$ metric and $\Phi$ a frame-dependent conformal
factor, $\Phi\in\det{T^*M}$. 
     
   A special class of conformal frames are those defined by a splitting
of the generalised tangent space $E$. A splitting is a map $TM\to
E$. It is equivalent to specifying a local two-form $B$ patched as $
   B_{(i)} = B_{(j)} - \dd\Lambda_{(ij)}$,
making $B$ a ``connection structure on a
gerbe''. 
This defines an isomorphism $E\simeq TM \oplus
T^*M$. If $\{\hat{e}_a\}$ is a generic basis for $TM$ and $\{e^a\}$ be
the dual  basis on $T^*M$, then together with some scaling function function $\phi$, we can define a
\emph{conformal split frame} $\{\hat{E}_A\}$ for $\tilde{E}$ by  
\begin{equation}
\label{eq:Csplit}
   \hat{E}_A = \begin{cases} 
           \hat{E}_a = \ee^{-2\phi}(\det e) \left(
              \hat{e}_a + i_{\hat{e}_a} B \right) 
               & \text{for $A=a$} \\
           E^a = \ee^{-2\phi}(\det e) e^a &  \text{for $A=a+10$}
         \end{cases} . 
\end{equation}
The class of conformal split frames defines a parabolic subgroup
$\Ggeom\times\bbR^+=(\GL(d,\bbR)\ltimes\bbR^{10\cdot9/2})\times\bbR^+\subset O(10,10)\times\bbR^+$. This reflects the fact that the patching elements
in the definition of $\tilde{E}$ lie only in this subgroup of
$O(10,10)\times\bbR^+$.

\paragraph{The Dorfman derivative and Generalised Connections}

The generalised tangent space 
admits a generalisation of the Lie derivative which encodes the
bosonic symmetries of the NSNS sector. Given
$V=v+\lambda\in\Gs{E}$, we define the Dorfman derivative~\cite{GMPW} $\Lgen_V$ acting on $W=w+\zeta\in \Gs{\tilde{E}}$ 
\begin{equation}
\label{eq:Lgen}
   \Lgen_V W = \mathcal{L}_v w + \mathcal{L}_v \zeta - i_w \dd\lambda .
\end{equation}

A \emph{generalised connection} is a 
first-order linear differential operator $\Dgen$ compatible with $O(10,10)\times\bbR^+$, such that, in frame indices, 
\begin{equation}
   \Dgen_M W^A = \der_M W^A + \Omega_M{}^A{}_B W^B - \Lambda_M W^A. 
\end{equation}
where $\Lambda$ is the $\bbR^+$ part of the connection and $  \Omega_M{}^{AB} = - \Omega_M{}^{BA}$ the $O(10,10)$ part.

Given a conventional connection $\nabla$ and a conformal split frame
of the form~\eqref{eq:Csplit}, we can construct the corresponding
generalised connection defined by lifting $\nabla$  to an action on $\tilde{E}$  
\begin{equation}
\label{eq:Dgen-embed}
   (\Dgen^\nabla_M W^A) \hat{E}_A= \begin{cases}
      (\nabla_\mu w^a) \hat{E}_a + (\nabla_\mu \zeta_a) E^a  
      & \text{for $M=\mu$} \\
      0  & \text{for $M=\mu+10$} 
      \end{cases} . 
\end{equation}

The \emph{generalised torsion} $T$ of a generalised connection
is defined in direct analogy to the conventional
definition. Let $\Lgen^\Dgen_V$ be the Dorfman
derivative~\eqref{eq:Lgen} with $\der$ replaced by $\Dgen$. The
generalised torsion is a linear map $T:\Gs{E}\to\Gs{\Lambda^2E\oplus\bbR}$. It is defined, for any $V\in\Gs{E}$, by 
\begin{equation}
T(V)\cdot W        = \Lgen^\Dgen_V W - \Lgen_V W .
\end{equation}
Viewed as a tensor, we find that $T \in \Lambda^3 E \oplus E $. As an example, we can calculate the torsion for the generalised connection
$\Dgen^\nabla$ defined in~\eqref{eq:Dgen-embed} when $\nabla$ has zero conventional torsion. Its generalised torsion is
\begin{equation}
\label{eq:Hphi-tor}
   T = - 4 H - 4 \, \dd \phi , 
\end{equation}
where $H = \dd B$.


\paragraph{Generalised metric and $O(9,1)\times O(1,9)$  structures}
\label{sec:OdOd}

Following closely~\cite{Gualtieri}, consider an $O(9,1)\times O(1,9)$ principal
sub-bundle of the $O(10,10) \times \bbR^+$ bundle. This is equivalent to specifying a conventional
metric $g$ of signature $(9,1)$, a $B$-field and a dilaton $\phi$. As such it clearly gives
the appropriate generalised structure to capture the NSNS supergravity
fields. We can write a generic $O(9,1)\times O(1,9)$ structure explicitly as 
\begin{equation}
\label{eq:OdOdexplicit}
\begin{aligned}
   \hat{E}^+_a &= \ee^{-2\phi}\sqrt{-g} 
      \left( \hat{e}^+_a + e^+_a + i_{\hat{e}^+_a}B \right) , \\
   \hat{E}^-_{\bar{a}} &= \ee^{-2\phi}\sqrt{-g} 
      \left( \hat{e}^-_{\bar{a}} - e^-_{\bar{a}} + i_{\hat{e}^-_{\bar{a}}}B \right) , 
\end{aligned}
\end{equation}
where $\{\hat{e}^+_a\}$ and $\{\hat{e}^-_{\bar{a}}\}$, and their
duals are two independent
orthonormal frames for the metric $g$. We see that the generalised tangent space splits
$
   E = C_+ \oplus C_-
$.

We can alternatively define the invariant \emph{generalised metric} 
\begin{equation}
 G = \big(\eta^{ab} \hat{E}^+_a\otimes\hat{E}^+_b + \eta^{\bar{a}\bar{b}} \hat{E}^-_{\bar{a}}\otimes\hat{E}^-_{\bar{b}} \big)
\end{equation}
and the fixed conformal density $\vol_G = \ee^{-2\phi}\sqrt{-g}$. By construction, $G$ parametrises the coset
$(O(10,10)\times\bbR^+)/O(9,1)\times O(1,9)$. Note that the infinitesimal bosonic symmetry transformations of supergravity are naturally encoded as the Dorfman derivative $\delta_V G = \Lgen_V G$.

\paragraph{Torsion-free, compatible connections}
\label{sec:genLC}

A generalised connection $\Dgen$ is compatible with the $O(9,1)\times
O(1,9)$ structure if $\Dgen G = 0 $. In analogy to the construction of the Levi--Civita connection, there always exists a torsion-free, compatible generalised connection $\Dgen$. However, it is not unique.

To construct such a connection we simply modify
$\Dgen^\nabla$. A generic generalised connection $\Dgen$ can always 
be written as $  \Dgen_M W^ A = \Dgen^\nabla_M W^A + \Sigma_M{}^A{}_B W^B $.
If $\Dgen$ is compatible and torsion-free, we find  
\begin{flalign*} 
   \Dgen_a w_+^b 
       &= \nabla_a w_+^b - \tfrac{1}{6}H_a{}^b{}_cw_+^c
           - \tfrac{2}{9}\big( 
              \delta_a{}^b \der_c\phi-\eta_{ac}\der^b\phi \big)w_+^c 
           + A^+_a{}^b{}_c w_+^c , &
   \Dgen_{\bar{a}} w_+^b 
       = \nabla_{\bar{a}} w_+^b - \tfrac{1}{2}H_{\bar{a}}{}^b{}_cw_+^c , \\ 
   \Dgen_{\bar{a}} w_-^{\bar{b}} 
       &= \nabla_{\bar{a}} w_-^{\bar{b}} 
           + \tfrac{1}{6}H_{\bar{a}}{}^{\bar{b}}{}_{\bar{c}}w_-^{\bar{c}}
           - \tfrac{2}{9}\big( 
              \delta_{\bar{a}}{}^{\bar{b}} \der_{\bar{c}}\phi
              - \eta_{\bar{a}\bar{c}}\der^{\bar{b}}\phi \big)w_-^{\bar{c}} 
           + A^-_{\bar{a}}{}^{\bar{b}}{}_{\bar{c}} w_-^{\bar{c}} ,
    &
  \Dgen_a w_-^{\bar{b}} 
       = \nabla_a w_-^{\bar{b}} 
           + \tfrac{1}{2}H_a{}^{\bar{b}}{}_{\bar{c}}w_-^{\bar{c}} ,
\end{flalign*}
where the undetermined tensors $A^\pm$ lie in $O(9,1)\times
O(1,9)$ representations that do not contribute to the torsion. We see explicitly that there is no unique torsion free compatible connection.

However, in supergravity we are interested in writing unambiguous expressions, so it is necessary to project out the $A^\pm$. Additionally, we will need to construct operators that act on spinors. Given that $M$ is spin, we can promote the local structure to $\Spin(9,1)\times Spin(1,9)$, so if $S(C_\pm)$ are then the spinor bundles associated to the sub-bundles
$C_\pm$, $\gamma^a$ and $\gamma^{\bar{a}}$ the corresponding gamma
matrices and $\epsilon^\pm \in \Gs{S(C_\pm)}$, we have that by
definition a generalised connection acts on spinors as    
\begin{equation}
   \Dgen_M \epsilon^+ = \der_M \epsilon^+ +
         \tfrac{1}{4}\Omega_M{}^{ab}\gamma_{ab}\epsilon^+ , \qquad
   \Dgen_M \epsilon^- = \der_M \epsilon^- +
         \tfrac{1}{4}\Omega_M{}^{\bar{a}\bar{b}}
            \gamma_{\bar{a}\bar{b}}\epsilon^- .
\end{equation}
There are four operators which can be built out of these derivatives
that are uniquely determined
\begin{equation}
\label{eq:Dgen-spin-uniq}
\begin{aligned}
   \Dgen_{\bar{a}}\epsilon^+ 
      &= \left( \nabla_{\bar{a}} 
           - \tfrac{1}{8}H_{\bar{a}bc}\gamma^{bc} \right) \epsilon^+ ,
           \qquad \gamma^a\Dgen_a\epsilon^+ 
      = \left( \gamma^a\nabla_a - \tfrac{1}{24}H_{abc}\gamma^{abc}
           - \gamma^a\der_a\phi \right) \epsilon^+ , \\
   \Dgen_a\epsilon^- 
   	&= \left( \nabla_a + \tfrac{1}{8}H_{a\bar{b}\bar{c}}
           \gamma^{\bar{b}\bar{c}} \right) \epsilon^- ,
           \qquad \gamma^{\bar{a}}\Dgen_{\bar{a}}\epsilon^- 
      = \left( \gamma^{\bar{a}}\nabla_{\bar{a}} 
           + \tfrac{1}{24}H_{\bar{a}\bar{b}\bar{c}}
               \gamma^{\bar{a}\bar{b}\bar{c}}
           -\gamma^{\bar{a}}\der_{\bar{a}}\phi 
               \right) \epsilon^- . 
\end{aligned}
\end{equation}
We can now also define measures of generalised curvature, a (traceless) \emph{generalised Ricci tensor} $\GenRic_{a\bar{b}}$ given by either 
\begin{equation}
\label{eq:GenRicSpin}
   \tfrac12 \GenRic_{a\bar{b}} \gamma^a \epsilon^+ 
      = \left[\gamma^a \Dgen_a , \Dgen_{\bar{b}} \right] \epsilon^+ ,\qquad
   \tfrac12 \GenRic_{\bar{a}b} \gamma^{\bar{a}} \epsilon^- 
      = \left[\gamma^{\bar{a}} \Dgen_{\bar{a}} , \Dgen_b \right] \epsilon^- ,
\end{equation}
and a \emph{generalised curvature scalar} $\GenS$
\begin{equation}
\label{eq:GenS+}
	-\tfrac14 \GenS \epsilon^+ = \big( \gamma^a \Dgen_a \gamma^b \Dgen_b 
	- \Dgen^{\bar{a}}\Dgen_{\bar{a}} \big) \epsilon^+ , \qquad
   -\tfrac14 \GenS \epsilon^- = \big( 
     \gamma^{\bar{a}} \Dgen_{\bar{a}} \gamma^{\bar{b}} \Dgen_{\bar{b}} 
     - \Dgen^{a}\Dgen_{a} \big) \epsilon^- .
\end{equation}
%

\paragraph{Fermionic and RR supergravity fields}
\label{sec:NSNS-fields}

The type II fermionic degrees of freedom fall into spinor and
vector-spinor representations of
$\Spin(9,1)\times\Spin(1,9)$. Note that since we are
in ten dimensions, we can further decompose $S(C_\pm)$ into spinor bundles
$S^\pm(C_+)$ and $S^\pm(C_-)$ of definite chirality under $\gamma^{(10)}$. The fermionic degrees of freedom then correspond to 
\begin{equation*}
   \psi^+_{\bar{a}} \in \Gs{C_-\otimes S^\mp(C_+)} , \quad
   \psi^-_a \in \Gs{C_+\otimes S^+(C_-)} , \quad 
   \rho^+ \in \Gs{S^\pm(C_+)} ,  \quad
   \rho^- \in \Gs{S^+(C_-)} , 
\end{equation*}
where the upper sign on the chirality refers to type IIA and the lower
to type IIB.

Let $\Gamma$ denote the $\Cliff(10,10;\bbR)$ gamma matrices, and $S^\pm_{(1/2)}$ the spin bundles of definite chirality under $\Gamma^{(20)}$. The RR field strengths $F = \sum_n F_{(n)}$ then transform as spinors of positive chirality for type IIA and negative for type IIB. Given the generalised metric structure, we have the decomposition
$\Cliff(10,10;\bbR)\simeq\Cliff(9,1;\bbR)\otimes\Cliff(1,9;\bbR)$
and hence we can identify $S_{(1/2)}\simeq S(C_+)\otimes S(C_-)$, allowing us to write $F$ in terms of $\Spin(9,1)\times\Spin(1,9)$ representations. Note that the self-duality conditions satisfied by $F$ become a chirality condition
under the operation 
\begin{equation}
\Gamma^{(-)} F = \tfrac{1}{10!} \epsilon^{\bar{a}_1 \dots \bar{a}_{10}}   \Gamma_{\bar{a}_1} \dots \Gamma_{\bar{a}_{10}} F = - F .
\end{equation}

Using the spinor norm on $S(C_-)$ we can equally well view $F \in
S_{(1/2)}$ as a map from sections of $S(C_-)$ to sections of $S(C_+)$. We
denote the image under this isomorphism as $
\bisp{F} : S(C_-) \rightarrow S(C_+) $ and the conjugate map, $\bisp{F}^T : S(C_+)
\rightarrow S(C_-)$.


\paragraph{Supersymmetry variations and equations of motion}
\label{sec:susy}

The expressions~\eqref{eq:Dgen-spin-uniq} describe precisely the fermionic supersymmetry transformations, allowing us to write them as
\begin{equation}
\begin{aligned}
   \delta \psi^+_{\bar{a}} &= \Dgen_{\bar{a}}\epsilon^+ + \tfrac{1}{16} 
\bisp{F}\gamma_{\bar{a}} \epsilon^- , 
	&\delta\rho^+ = \gamma^a\Dgen_a\epsilon^+ , \\
   \delta \psi^-_a &= \Dgen_a\epsilon^- + \tfrac{1}{16} \bisp{F}^T \gamma_{a} \epsilon^+, 
    	&\delta\rho^- = \gamma^{\bar{a}}\Dgen_{\bar{a}}\epsilon^- . 
\end{aligned}
\end{equation}

For the NSNS bosonic fields, we have
\begin{equation}
   \delta G_{a\bar{a}} = \delta G_{\bar{a} a} 
      = 2 \left( \bar{\epsilon}^+ \gamma_a \psi^+_{\bar{a}} 
         + \bar{\epsilon}^- \gamma_{\bar{a}} \psi^-_a \right) \vol_G^2 + 2 \left(  \bar{\epsilon}^+ \rho^+ + \bar{\epsilon}^- \rho^-  \right)G_{a\bar{a}},
\end{equation}
and the variation of the RR potential $A$ can be written as a bispinor
\begin{align}
   \tfrac{1}{16} (\delta \bisp{A}) 
      = \big( \gamma^a \epsilon^+ \bar\psi^-_a
         - \rho^+\bar\epsilon^- \big) 
         \mp \big( \psi^+_{\bar{a}}\bar\epsilon^- \gamma^{\bar{a}}
         + \epsilon^+\bar\rho^- \big) ,
\end{align}
where the upper sign is for type IIA and the lower for type IIB.

Finally, we rewrite the supergravity equations of
motion. From the generalised Ricci tensor and scalar, we find that
the equations of motion for $g$, $B$ and $\phi$ can be written as the analogue of the Einstein equations
\begin{equation}
\label{eq:G-eom}
   \GenRic_{a\bar{b}} 
      = - \tfrac{1}{16}\vol_G^{-1}\mukai{F}{\Gamma_{a \bar{b}} F}, \qquad \GenS = 	0 ,
\end{equation}
where we have made use of the Mukai pairing between two spinors~\cite{GCY}. The equation of motion for the RR fields has the form
\begin{equation}
\label{eq:F-eom}
	\tfrac12 \Gamma^A \Dgen_A F = \dd F = 0 ,
\end{equation}
and the full bosonic pseudo-action is simply given by
\begin{equation}
\label{eq:SB-gen}
\begin{aligned}
   S_B = \frac{1}{2\kappa^2}\int \left(  \vol_G  \GenS 
            + \tfrac{1}{4} \mukai{F}{\Gamma^{(-)} F} \right) .
\end{aligned}
\end{equation}
Note that the Mukai pairing is a top-form
which can be directly integrated. 

The fermionic action is
\begin{equation}
\begin{aligned}
   S_F = -\frac{1}{2\kappa^2}\int &2 \vol_G \Big[ 
         \bar\psi^{+\bar{a}} \gamma^b \Dgen_b \psi^+_{\bar{a}}
         + \bar\psi^{-a} \gamma^{\bar{b}} \Dgen_{\bar{b}} \psi^-_{a}  
         + 2 \bar\rho^+ \Dgen_{\bar{a}} \psi^{+\bar{a}}
         + 2 \bar\rho^- \Dgen_a \psi^{-a} \\
         &- \bar\rho^+ \gamma^a \Dgen_a \rho^+ 
         - \bar\rho^- \gamma^{\bar{a}} \Dgen_{\bar{a}} \rho^-  -\tfrac{1}{8} \Big( \bar\rho^+ \bisp{F}\rho^-
         + \bar\psi^+_{\bar{a}} \gamma^a \bisp{F}\gamma^{\bar{a}} \psi^-_a \Big)
         \Big].
\end{aligned}
\end{equation}

We have thus rewritten type II supergravity using a torsion-free compatible connection $\Dgen$, in direct analogy to conventional gravity. As a result, the theory has manifest $\Spin(9,1)\times\Spin(1,9)$ covariance together with an extension of the diffeomorphism group by the $B$-field gauge transformations. In~\cite{CSW2} we show that the same formalism applies to 11-dimensional supergravity restricted to $d<8$ dimensions.



\paragraph{Acknowledgments} A.C. would like to thank the organisers of  the ``XVII European Workshop in String Theory'' for the opportunity to present this work. C.~S-C.~is supported by an STFC PhD studentship. A.~C.~is
supported by the Portuguese FCT under grant SFRH/BD/43249/2008.


\end{document}